\documentclass{IEEEcsmag}

\usepackage[colorlinks,urlcolor=blue,linkcolor=blue,citecolor=blue]{hyperref}
\expandafter\def\expandafter\UrlBreaks\expandafter{\UrlBreaks\do\/\do\*\do\-\do\~\do\'\do\"\do\-}
\usepackage{upmath}

\usepackage{tablefootnote}
\usepackage{framed}
\usepackage{tabularx}
\usepackage{booktabs}
\usepackage[most]{tcolorbox}
\usepackage{xcolor}
\usepackage{tikz}
\usetikzlibrary{positioning}

\definecolor{lightgray}{HTML}{F5F5F5}
\definecolor{lightblue}{HTML}{AFCBDB}
\definecolor{orange}{HTML}{F7A34B}

\definecolor{SectionOrange}{HTML}{EEA740}

\newcommand{\pquote}[3]{\textcolor{#3}{\textbf{#1 (#2):} }}

\newcommand{\dialoguegpt}[2]{
    \begin{coloredframe}{#1}
    \vspace{5px}
    \footnotesize 
    \begin{zeroindent} #2 \end{zeroindent}
    \vspace{5px}
    \end{coloredframe}
}

\definecolor{quotecolor}{HTML}{204985}
\newenvironment{zeroindent}
  {\par\setlength{\parindent}{0pt}}
  {\par}

\newcommand{\tracked}[1]{%
  \begingroup
  \def\reviewerColor{}%
  \textcolor{black}{#1}%
  \endgroup
}

\newcommand{\minor}[1]{%
  \begingroup
  \def\reviewerColor{}%
  \textcolor{black}{#1}%
  \endgroup
}

\newcommand{\casebox}[3]{%
\noindent
\begin{tikzpicture}

\node[inner sep=1mm, anchor=north west] (container) at (0, 0) {
    \begin{tcolorbox}[
        colback=lightblue, colframe=black!0,
        boxrule=0pt, width=7.5cm, arc=0mm,
        left=4mm, right=4mm, top=2mm, bottom=2mm,
        boxsep=0pt
    ]
    \centering\bfseries #1: #2
    \end{tcolorbox}
};

\node[anchor=north west] at ([yshift=2.2mm]container.south west) {
    \begin{tcolorbox}[
        colback=lightgray, colframe=black!0, 
        boxrule=0pt, width=7.5cm, arc=0mm,
        left=4mm, right=4mm, top=2mm, bottom=2mm,
        boxsep=0pt
    ]
    \textbf{Policy Focus:}
    #3
    \end{tcolorbox}
};

\end{tikzpicture}%
}

\usepackage[framemethod=TikZ]{mdframed}
\definecolor{backgroundpage}{HTML}{204985}

\newenvironment{coloredframe}[2][]{
    \mdfsetup{
        skipabove=2pt, 
        hidealllines=true, leftline=true,      
        innerlinewidth=3pt, innerlinecolor=#2, 
        linewidth=0pt,
        backgroundcolor=#2!5
    }
    \begin{mdframed}}
    {\end{mdframed}}
\jvol{XX}
\jnum{XX}
\paper{8}
\jname{Special Issue on AIware in the FM Era}
\jtitle{LLM Company Policies and Policy Implications
in Software Organizations}

\begin{document}

\sptitle{Special Issue on AIware in the FM Era}

\title{LLM Company Policies and Policy Implications in Software Organizations}

\author{Ranim Khojah}
\affil{Chalmers University of Technology and University of Gothenburg, Sweden}

\author{Mazen Mohamad}
\affil{RISE Research Institutes of Sweden and Chalmers University of Technology, Sweden}

\author{{L}inda Erlenhov}
\affil{Chalmers University of Technology and University of Gothenburg, Sweden}

\author{Francisco Gomes de Oliveira Neto}
\affil{Chalmers University of Technology and University of Gothenburg, Sweden}

\author{Philipp Leitner}
\affil{Chalmers University of Technology and University of Gothenburg, Sweden}


\begin{abstract}\looseness-1
The risks associated with adopting large language model (LLM) chatbots in software organizations highlight the need for clear policies. We examine how 11 companies create these policies and the factors that influence them, aiming to help managers safely integrate chatbots into development workflows.
\end{abstract}

\maketitle

\chapteri{A}rtificial Intelligence (AI) has revolutionized the toolbox of software engineers, allowing them to automate software creation, receive insightful recommendations, and perform a wide range of tasks~\cite{khojah2024beyond}.
\tracked{In software organizations, the software product is gradually evolving to AI-powered software (AIware) with the use of AI, more specifically, large language models (LLMs) in the development process \cite{hassan2024fm}. LLMs are increasingly seen as valuable tools for improving productivity, which motivated enterprises to adopt them \cite{sharma2024sme}.}

However, these models have introduced risks and concerns that impact the organization, the software engineers, and the product. \tracked{Integrating LLMs into software development raises challenges related to the quality and ownership of generated content \cite{ozkaya2023applications}, which complicates accountability and can affect product reliability. In addition, interactions with LLMs (e.g., through external APIs) may expose organizations to liability where developers unintentionally transmit sensitive data, resulting in legal repercussions \cite{herbold2025legal}. This risk can be amplified when developers use chat-based interfaces, which may create a false sense of human-like familiarity that obscures security and privacy concerns, ultimately impacting trust \cite{nahar2024beyond}. }

These \tracked{concerns} lead companies to either reject LLM adoption in development or implement policies to constrain their use.
While policies are an effective way to mitigate these risks, the rapid evolution of these technologies means that defining clear rules and boundaries to guide their use in software engineering is urgent and unavoidable, but also highly challenging.
To better understand what \minor{software engineers}, managers, and decision makers need to consider when adopting LLM chatbots, we interview practitioners in management roles from 11 software organizations across four countries in Europe and Asia about their LLM policy.

\section{Data Collection and Methodology}

\begin{table*}[!ht]
\centering
\scriptsize
\caption{Information about our interviewees, their companies, the chatbots they use, whether they are used in software development activities, \protect\tracked{and the corresponding contexts (explained in Section ``What LLM Chatbot Policies Cover'')}. Company sizes were classified according to the categories recommended by the European Commission~\cite{euro-org-size}.}
\label{tab:participants}
\begin{tabularx}{\linewidth}{llllllll}
\toprule
 \textbf{Role} & 
\textbf{Company} & \textbf{Domain} & \textbf{Size} &  \textbf{Country} & \textbf{Chatbot types} & \textbf{Use in dev.} & \textcolor{black}{\textbf{Context}}\\
\midrule
 CEO &  EduCo     & Education            & Micro    & Netherlands & Closed-source (license) & Yes & \textcolor{black}{C} \\
Founder &  ReaderCo  & Reading tech & Micro   & Sweden   & Closed-source (no license) &  Yes  & \textcolor{black}{B} \\
 Team Lead & CloudCo   & Cloud platform       & SME    & Switzerland & Closed-source (no license) &  Yes & \textcolor{black}{B}\\
 CTO &  EyeCo     & Eye-tracking tech    & SME    & Sweden   & Closed-source (no license) &   Yes & \textcolor{black}{B} \\
 Manager & SysManCo  & Systems mgmnt.   & SME    & Sweden   & Closed-source (license) &   Yes & \textcolor{black}{C} \\
Manager & TestCo    & Testing consulting  & SME    & Sweden    & Depends on customer &  Yes  & \textcolor{black}{B,C,D}\\
Team Lead & ConsultCo & Software consulting & Large  & Sweden & Closed-source (license) &  Yes & \textcolor{black}{C} \\
Team Lead &  AeroCo    & Aviation             & Large    & Sweden   & Open-source (local) & No  & \textcolor{black}{A,D} \\
Product Owner & FlightCo  & Aviation             & Large & Sweden   & Closed-source (license) &  No & \textcolor{black}{A,C} \\
Manager & AutoCo    & Automotive           & Large    & Sweden    & Closed-source (license) & No  & \textcolor{black}{A,C}\\
Manager &  ITServeCo & IT services          & Large & India    & Open-source (local)  & Yes   & \textcolor{black}{C,D} \\
& & & & & and Closed-source (license) & & \\
\bottomrule 
\\
\end{tabularx}
\end{table*}

\begin{coloredframe}{SectionOrange}

We spoke with 11 managers across 11 different organizations \tracked{that use or have access to LLM chatbots (Table~\ref{tab:participants}). The selection of managers was based on their authority to make or influence decisions regarding the use of LLM-based chatbots in software development at their organizations. These chatbots fall into three categories: (i) self-hosted open-source models, and commercial closed-source ones with either (ii) free subscriptions or (iii) enterprise licenses}.

Our participants held a range of roles---including team leads, process managers, and CTOs---which helped us capture diverse perspectives on how LLM chatbot adoption affects software engineering processes, teams, products, and customer value. Most of our participants work in Sweden, but we have also interviewed participants in Switzerland, India, and the Netherlands. 

The interviews were semi-structured, each lasting between 30 and 60 minutes. Participants consented after being informed about the study's purpose, anonymization, interview recording, and their right to opt out.
We analyzed the transcripts using thematic analysis. 
\tracked{In an initial coding round, three researchers independently coded three interviews, each researcher overlapping with another on two interviews to measure consistency. \minor{A total of 73\% of excerpts} (166/225) overlapped between all researchers, indicating agreement in coding, with most discrepancies involving excerpts where chatbots were used outside of development contexts.}

Following this, all authors engaged in three collaborative sorting sessions. We reviewed and refined the codes, ultimately grouping them into broader themes. After three rounds, we reached thematic saturation, identifying key issues such as organizational change and the creation of policies to guide LLM chatbot use. The interview protocol, list of extracted codes, and themes are available in our reproduction package to support future researchers.\tracked{\footnote{\url{https://doi.org/10.5281/zenodo.15173862}}}

\end{coloredframe}

\section{Why LLM Policies Are Needed}

In many organizations, the push towards LLMs has been bottom-up rather than top-down \tracked{\cite{lambiase2025cultural}}, in the form of a grassroots movement of developers experimenting with LLMs and chatbots to automate repetitive tasks \tracked{and build software more efficiently}. 
However, there are risks to such unstructured adoption of LLMs by individual engineers, namely the lack of a clear plan and rules of engagement in the form of a clear \emph{policy}.

During our interviews, managers expressed the concern that less experienced practitioners might not always critically assess the output generated by LLMs. This can lead to challenges in ensuring the quality and reliability of the product, and may affect the level of trust and confidence between management and engineering teams.

\dialoguegpt{quotecolor}{
\pquote{CEO}{EduCo}{quotecolor} ``We noticed [that] some juniors took [the chatbot answer] too much for granted''
}

\minor{Moreover}, managers almost unanimously raised the risk of software engineers unintentionally exposing intellectual property (IP) or sensitive customer data in their prompts to LLMs. Such incidents could violate data protection regulations like GDPR, leading to legal penalties, loss of customer trust, and lasting damage to the company's reputation and financial stability.

\dialoguegpt{quotecolor}{
\pquote{Founder}{ReaderCo}{quotecolor} ``We have been afraid from the start that proprietary algorithms and stuff that we implement [are] indexed and used by chatbots''
}

Therefore, in response to the various risks raised, \tracked{the interviewed managers recognized the need to establish policies, and shared some of their approaches and steps in formulating and enforcing such policies.}

\section{How Industrial LLM Policies Work} 

\begin{figure*}[!ht]
    \centering
    \includegraphics[width=\linewidth]{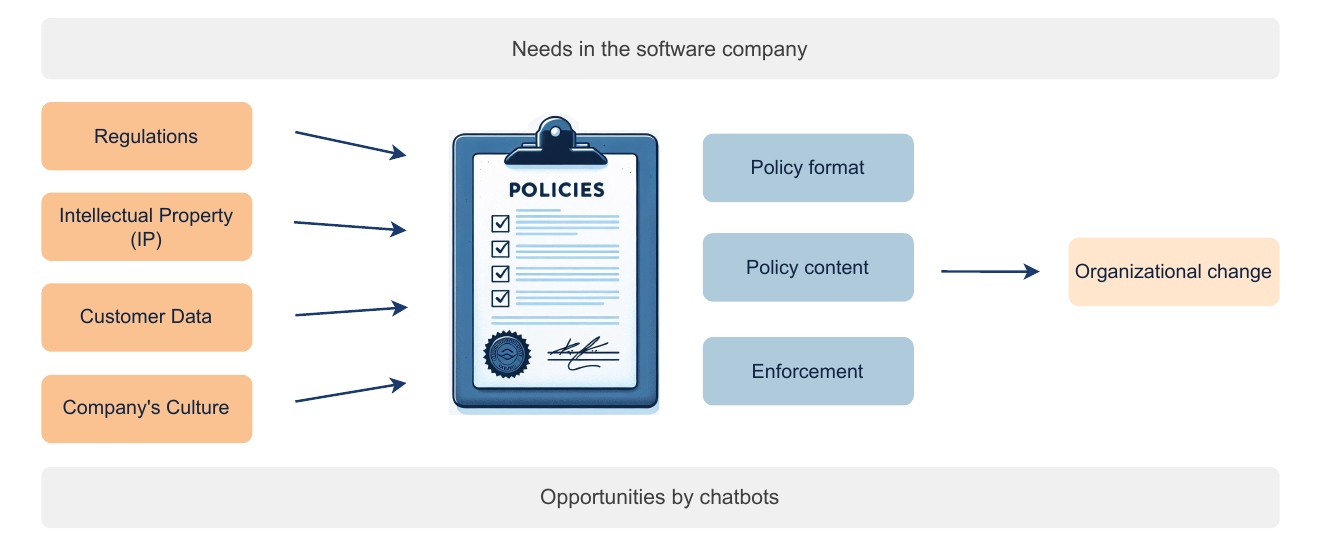}
    \caption{Factors that impact \protect\tracked{AI} policy creation in software organizations. \protect\minor{Note that the figure shows how the policies contribute to organizational change. While other factors may also lead to change, we focus on the specific role of policy within that broader context.}}
    \label{fig:policy-factors}
\end{figure*}

We found that policies are framed by (i) the needs of the company (e.g., minimize risks of compromising sensitive data), and (ii) the opportunities chatbots offer, such as boosting development team productivity. \tracked{Our interviews revealed varied factors influencing policy creation, communication, and enforcement. While not necessarily exhaustive, these factors (Figure~\ref{fig:policy-factors}) reflect insights on the main components our participants identified as critical to shaping chatbot policies.}

At the same time, policies are not just static documents. They are expected to influence ways of working, triggering organizational changes that, in turn, open up new needs and opportunities for chatbot use in software engineering, such as new roles within the organization. We discuss such organizational changes in the ``Preparing Companies for an LLM Era'' section.

\subsection{Policy Drivers}

Compliance with \textbf{regulations} and industry standards is a key driver of policy creation. Several participants specifically mentioned the EU AI Act\footnote{\url{https://artificialintelligenceact.eu}}, which requires developers to clearly label LLM-generated and manipulated content. In the absence of chatbot-specific standards, three companies relied on ISO 27001\tracked{\footnote{\url{https://www.iso.org/standard/27001}}} for information security that, among other requirements, demands classifying the data into sensitivity levels.

\dialoguegpt{quotecolor}{
\pquote{Manager}{SysManCo}{quotecolor} ``We did ISO 27001 and the big part of that work was to label the data (...) for what is sensitive and not sensitive (public, internal, confidential, critical data).''
}

We quickly found that the term ``data'' was too broad and we needed to identify more specific levels of meaning to capture the nuances in how organizations assess what information can or cannot be shared with chatbots. For instance, some policies focused on protecting \textbf{intellectual property} (IP) by prohibiting developers from including code or requirements specifications in chatbot prompts. 

\dialoguegpt{quotecolor}{
\pquote{Team Lead}{AeroCo}{quotecolor} ``Our value is in our source code, so we try to keep that a secret.''
}


However, for some companies, IP is not the source code, but rather custom data they have collected or acquired. In these cases, the policy is more open to allowing sharing the code with the chatbots, but more restrictive with regard to sharing data.

\dialoguegpt{quotecolor}{
\pquote{CTO}{EyeCo}{quotecolor} ``Cursor indexes quite a large part of your code base, because then it works better because it has a larger context, and that is OK for us.''
}

Regardless of the company's size or domain, our interviewees were in agreement that \textbf{customer data} needs to be treated with utmost care, and is always considered sensitive.



Finally, we have observed that the size and domain of a company significantly influence its \textbf{company culture}, and consequently, the strictness of its policies regarding LLM usage. For example, micro and small companies that rely heavily on open-source code may not have a formal policy in place and trust their employees to use any chatbot, while larger companies or those in heavily regulated \tracked{industries} implement much stricter guidelines, going as far as blocking \tracked{AI} tools through their firewall.

\subsection{Policy Creation and Enforcement}

Key steps when creating a policy include (1)
defining rules and guidelines to include, (2) deciding on the format for communicating the policy, and (3) a plan on how to enforce it. The content of the policy reflects important facets tied to the specific context in which the organization operates. Therefore, in this section, we focus on policy format and enforcement, 
while we discuss their content later.

The \textit{policy format} varies across companies. Three participants reported using a formal policy document, while another company integrated the policy directly into the chatbot, including it in the terms and conditions shown in the chatbot interface.
In most cases, however, companies (only) communicated their policies verbally, by email, or through company-wide announcements. In such cases, the policies were concise and often reduced to a simple guideline, such as ``do not share confidential information''.
Participants in smaller companies noted that this approach worked well and had not led to misunderstandings. \minor{EduCo suggests that this} is mostly due to smaller team sizes and shorter communication channels that allow for direct information sharing.
\tracked{Another} reason for this informal approach is that drafting formal policies requires upfront research and, ideally, a legal team --- a resource not always available to smaller companies. Two of the large companies we interviewed mentioned that their legal team handled the policy creation, or that they outsourced the research and policy drafting to consulting firms.

\dialoguegpt{quotecolor}{
\pquote{Manager}{ReaderCo}{quotecolor} ``We don't have enough lawyers to speak to (...) or a legal team to get through this.''
}

One challenge recognized by two companies was that if the policy is documented and stored elsewhere (e.g., on an intranet or shared drive), developers are likely to overlook it. To address this, they integrated the policy directly into the chatbot interface for them to read \minor{before being granted} access to the chatbot.

To promote employee compliance and ensure they stay up to date with policy changes, participants discussed different mechanisms for \textit{enforcement}. The most common approach was to hold training sessions where software engineers were taught how to apply the policy in practice and what types of use were prohibited, \tracked{but also} how to \tracked{use} chatbots \tracked{effectively}. Managers generally found these trainings more effective than simply referring employees to the policy document.

\dialoguegpt{quotecolor}{
\pquote{Manager}{ITServeCo}{quotecolor} ``[The policy] is being circulated within the team members and every six months we reconduct the training (...) for all new employees and juniors.''
}

\tracked{Policy} training needs to go beyond traditional security training. It must also focus on responsible use, helping employees recognize potential risks such as bias or misinformation, and developing skills such as prompt engineering. \tracked{At ITServeCo, one successful strategy was to conduct these trainings every six months and tailor the content to specific roles. For example, policy training for software testers focused on cases involving acceptance criteria and user stories.}

Depending on how strict a policy item is, companies take different measures to enforce compliance. For example, ITServeCo prohibits using chatbots other than the ones they provide. To ensure compliance, they implemented technical controls such as firewalls and blocked access to external chatbot services. Another strategy reported is to take a lighter approach by encouraging employees to experiment with different technologies, offering special subscriptions to those who need access to external chatbots.

Some companies faced a more challenging scenario when it came to enforcing policies. They opted \textit{not} to actively promote the use of chatbots in software development activities. However, they acknowledged that, even with clear restrictions, they had limited control over which tools engineers might choose to use independently. To address this gap while maintaining flexibility, two companies (AutoCo and AeroCo) offered safer alternatives, such as a licensed closed-source solution or locally hosted chatbots. Although these options were made available to employees, they were neither explicitly recommended nor heavily promoted, largely due to high costs.

\tracked{In general, we observed a trade-off between technical enforcement mechanisms (e.g., network restrictions and feature disabling) and other organizational strategies (e.g., offering alternatives and using trust-based models). While hard controls like firewalls are more enforceable, they may hinder innovation or lead to workarounds. In contrast, softer strategies provide flexibility but depend heavily on culture and employee cooperation with management.}

\minor{In particular, the culture of the company} plays a significant role in shaping how policies are created, communicated, and enforced. A culture that values open-mindedness and continuous improvement enables the company to learn from experience and adjust its policies or enforcement strategies as needed. The CEO of EduCo summarizes this perspective well when reflecting on areas where their own policy can improve.

\dialoguegpt{quotecolor}{
\pquote{CEO}{EduCo}{quotecolor} ``We see mistakes more as a learning opportunity for everyone; (...) there was also something wrong in our training or policies.''
}


The contrasting strategies we observed highlight how both policy format and enforcement must be tailored to fit the organization's culture and context. 
Standardized enforcement measures risk \minor{clashing with the organization's values} and practices, reducing their effectiveness.

\section{What LLM Chatbot Policies Cover} 

Regarding \textit{policy content}, we observe a consistent presence of key elements in the policies of our participating companies, such as the restrictions on shared data. 
\tracked{However, the companies differed in the types of chatbots they use and in whether these chatbots are permitted for development-related tasks, such as code generation. These differences influenced which policy elements were prioritized and focused on by our participants}. 
\tracked{We group these differences into four usage and policy \textbf{contexts}: A) Non-development usage only, B) Unlicensed closed-source model, C) Closed-source model with enterprise license, D) Self-hosted open-source model (See Table \ref{tab:participants})}.
By highlighting \tracked{their} focus, managers can better align the company's goals (e.g., complying with regulatory requirements or encouraging experimentation with new technologies) and more easily decide which specific rules to include in the policy document. Note that these \tracked{contexts are \textit{not} mutually exclusive; a single policy might cater to multiple contexts, with varying levels of detail, and a company may use several chatbots covered under the same policy. We aim to empower decision makers by raising awareness of these contexts, especially when adjusting current development practices.}

\casebox{Context A}{Non-Development Usage Only}{Define permitted and prohibited use cases for AI chatbots, explicitly disallowing development-related tasks.}

\tracked{Software organizations within safety-critical domains (e.g., automotive and aviation) centered their policy on how chatbots should and should not be used. In such cases, companies prohibit chatbot use to generate code contributions or restrict it to non-development tasks such as writing emails. Although software development involves many activities, we observed that our participants were cautious about allowing chatbots to assist with coding tasks in particular.}

\dialoguegpt{quotecolor}{
\pquote{Manager}{AutoCo}{quotecolor} ``We don't generate code or use [chatbots] in our code base at all.''
}

\tracked{To ensure these policies are followed, companies need access to the interactions between employees and chatbots. Therefore, they prefer either hosting their own chatbot on internal servers or using commercial chatbots that provide licensed access to employees' prompts. Although they do not actively monitor the prompts' compliance with the policy, they want to store this data in case it is needed.}
\\

\casebox{Context B}{Unlicensed Closed-Source Model}{Restrict the types of data that can be shared to mitigate privacy and security risks.}


\tracked{Another context that we observed is when companies allowed the use of commercial chatbots (e.g., ChatGPT) in the development \textit{without} purchasing a license. The free subscription (no license) of closed-source chatbots often allows providers to reuse interaction data to retrain their models, making this restriction critical for managers. As a result, we observed that the corresponding LLM policies focus mainly on the type of data that is not allowed to be shared with the chatbot. These policies emphasized the need to prohibit sharing confidential information such as customer data or intellectual property.}

Those restrictions also implicitly define how chatbots can be used. For example, if sharing production code is prohibited, activities like code repair or refactoring may not be feasible, whereas code generation could still be allowed. 
\tracked{While stricter in companies without a license, such restrictions appeared in all policies (though in varying levels of detail), which reflects caution around legal, financial, and reputation risks, including potential impact on customer trust.}

\dialoguegpt{quotecolor}{
\pquote{Manager}{SysManCo}{quotecolor} ``Big fine if you misuse [chatbot], it might hit quite bad and the company reputation gets affected (...) If you get sued because of misuse of a certain technology, there is definitely financial impact.''
}

\casebox{Context C}{Closed-Source Model With Enterprise License}{Specify approved chatbots and provide structured guidelines for access and setup.}

\tracked{In companies using enterprise-licensed chatbots, policies often focus on restricting employees to using only the chatbots officially provided by the company. Specific requirements for the chatbot configurations are also commonly provided\tracked{, e.g.}, disabling the option to share data with the chatbot providers for further training, or requiring employees to authenticate and access the chatbot through an internal portal.}

\dialoguegpt{quotecolor}{
\pquote{CEO}{EduCo}{quotecolor} ``We have to turn off the checkbox that the data you input can be used to train the model.''
}

\tracked{These policy measures allow the company to create a controlled environment that enables safe chatbot use, even without the full control they would have by hosting their own chatbot, which can be costly to implement and maintain for startups and SMEs.}
\\

\casebox{Context D}{Self-hosted Open-source Model}{Emphasize internal responsibility for verifying the correctness and quality of the model's output.}

In cases where chatbots are hosted locally, companies can more easily enforce requirements (e.g., not sharing a specific type of data) through system design rather than relying on users to comply manually. For instance, AeroCo built a custom chatbot with multiple environments, each tailored to different users' security clearances. Developers, for instance, access a restricted environment that prevents uploading or analyzing documents, while managers with higher privileges can use these features.

Another aspect that emerged, although not always formally stated in policies, was authority over chatbot usage. In some companies, junior engineers were initially restricted from using chatbots until they demonstrated responsible behavior and earned trust to access them independently.

When using chatbots for development tasks, engineers must ensure that the generated output is used safely to avoid harming the product under development. Most companies emphasized the need for code review and verification whenever an artifact (e.g., code, requirements, test cases) is produced by an LLM chatbot. \tracked{While all participants recognize the importance of verification, companies hosting models locally stressed it even more and noted that open-source models typically underperform in code generation. 
This is also evident in modern code-reasoning benchmarks where the top results are often held by closed-source LLMs, with the best open-source models improving but still below closed-source ones~\footnote{\url{https://www.swebench.com}}.}
Currently, chatbot-generated results are verified using the same process as other artifacts created by an engineer or from a third party, such as testing or manual inspection. 

\dialoguegpt{quotecolor}{
\pquote{Team Lead}{AeroCo}{quotecolor} ``If we introduce a third-party component or third-party code that we haven't written ourselves, it needs to go through quite rigorous testing before we can use it.''
}

Some companies pointed out that further research is needed to define additional verification steps tailored to chatbot output, which could then be incorporated into policies and help assure customers that the final product meets quality standards and delivers value.

\subsection{\protect\minor{LLM Policy Gaps}}
\minor{Two interesting gaps we noted were that} none of the companies we interviewed addressed accountability (e.g., what happens if an employee violates the policy) or copyright concerns in their policies (e.g., how to ensure chatbot output does not rely on copyrighted content). We speculate this is partly because chatbot-related policies are still new and evolving. Interviewees explained that these topics were not included simply because such situations have not yet occurred.

Less attention to copyright concerns was also recently linked to loopholes in general AI regulations, such as the EU AI Act, which many companies rely on\footnote{\url{https://www.theguardian.com/technology/2025/feb/19/eu-accused-of-leaving-devastating-copyright-loophole-in-ai-act}}. These loopholes appear to allow the use of AI-generated content even if it is trained on copyrighted data. For now, the legal situation remains unclear, and companies expect to update their policies as future court rulings and regulations emerge.
\\
\\





\section{Preparing Companies for an LLM Era} 

The adoption of LLM chatbot policies is already driving organizational changes in software companies. Software process models such as agile include several activities like daily stand-ups or sprint retrospectives that strengthen team communication. These activities will become even more important with chatbot adoption, helping to maintain team bonding and support informal knowledge sharing that chatbots cannot replace. Meanwhile, new activities related to chatbot governance emerge, such as monitoring prompts, promoting responsible use, and tracing LLM-generated artifacts. \tracked{Those practices help enforce policies and compliance, which is only feasible when the chatbot is under company control, such as in local or enterprise-licensed chatbots
, e.g., ChatGPT Enterprise.\footnote{\url{https://help.openai.com/en/articles/10875114-user-analytics-for-chatgpt-enterprise-and-edu-public-beta}}}

\tracked{Currently, auditing chatbot usage often falls on managers, but ultimately, a new role for chatbot governance may be required.}
Policies allowing chatbot-assisted code generation will also reshape developer roles, \tracked{potentially} shifting the focus from writing code to verifying and composing it~\cite{lyu2024automatic}. Unlike reviewing a single merge request, developers might need to evaluate multiple LLM-generated solutions across different prompts. 
Interestingly, despite public debate about chatbots replacing software engineers, our interviews suggest the opposite: chatbots create greater demand for engineers, though equipped with new skills~\cite{nahar2024beyond}.

\dialoguegpt{quotecolor}{
\pquote{Manager}{ITServeCo}{quotecolor} ``We have adapted existing roles [of engineers], but we are adding members to our team to take care of the work that these [engineers] are doing. (...) AI has added members to the team since we have a lot of work going on because of the updates and other steps.''
}

To support this shift, companies are starting to design specialized training aligned with their chatbot policies and practitioner roles (e.g., testers, developers, managers). At ITServeCo, targeted training programs have already improved employees' perceived productivity and efficiency. 

Through interviews with managers across a variety of companies, domains, and roles, we captured a broad view of
\tracked{the} key factors and priority areas that shape \tracked{LLM} policies in industry.
\tracked{Managers and decision makers can draft policy documents that reflect their own company's context and culture, guided by real-world practices of the companies we interviewed.}
\tracked{Documenting such LLM policies helps companies} navigate the era of AI-driven tools with greater clarity and avoid becoming overwhelmed when faced with important decisions regarding LLMs and software development.

\section{ACKNOWLEDGMENTS}
This work was partially supported by the Wallenberg AI, Autonomous Systems and Software Program (WASP) funded by the Knut and Alice Wallenberg Foundation.

\def\refname{REFERENCES}

\bibliographystyle{ieeetr}
\bibliography{bib}

\begin{thebibliography}{1}

\bibitem{khojah2024beyond}
R.~Khojah, M.~Mohamad, P.~Leitner, and F.~G. de~Oliveira~Neto, ``Beyond code generation: An observational study of chatgpt usage in software engineering practice,'' in {\em Proceedings of the 32nd ACM International Conference on the Foundations of Software Engineering}, vol.~1 of {\em FSE 2024}, (New York, NY, USA), Association for Computing Machinery, July 2024.

\bibitem{hassan2024fm}
A.~E. Hassan, D.~Lin, G.~K. Rajbahadur, K.~Gallaba, F.~R. Cogo, B.~Chen, H.~Zhang, K.~Thangarajah, G.~Oliva, J.~J. Lin, W.~M. Abdullah, and Z.~M.~J. Jiang, ``{Rethinking Software Engineering in the Era of Foundation Models: A Curated Catalogue of Challenges in the Development of Trustworthy FMware},'' in {\em Companion Proceedings of the 32nd ACM International Conference on the Foundations of Software Engineering}, FSE 2024, (New York, NY, USA), p.~294–305, Association for Computing Machinery, 2024.

\bibitem{sharma2024sme}
S.~Sharma, G.~Singh, N.~Islam, and A.~Dhir, ``{Why Do SMEs Adopt Artificial Intelligence-Based Chatbots?},'' {\em IEEE Transactions on Engineering Management}, vol.~71, pp.~1773--1786, 2024.

\bibitem{ozkaya2023applications}
I.~Ozkaya, ``Application of large language models to software engineering tasks: Opportunities, risks, and implications,'' {\em IEEE Software}, vol.~40, no.~3, pp.~4--8, 2023.

\bibitem{herbold2025legal}
S.~Herbold, B.~Valerius, A.~Mojica-Hanke, I.~Lex, and J.~Mittel, ``{Legal Aspects for Software Developers Interested in Generative AI Applications},'' {\em IEEE Software}, vol.~42, no.~2, pp.~68--75, 2025.

\bibitem{nahar2024beyond}
N.~Nahar, C.~K{\"a}stner, J.~Butler, C.~Parnin, T.~Zimmermann, and C.~Bird, ``{Beyond the Comfort Zone: Emerging Solutions to Overcome Challenges in Integrating LLMs into Software Products},'' in {\em Proceedings of the 47th International Conference on Software Engineering: Software Engineering in Practice}, ICSE-SEIP '25, 2025.

\bibitem{euro-org-size}
E.~Commision, ``{Internal Market, Industry, Entrepreneurship and SMEs},'' 2021.
\newblock Accessed on April 6, 2025.

\bibitem{lambiase2025cultural}
S.~Lambiase, G.~Catolino, F.~Palomba, F.~Ferrucci, and D.~Russo, ``Investigating the role of cultural values in adopting large language models for software engineering,'' {\em ACM Trans. Softw. Eng. Methodol.}, Mar. 2025.

\bibitem{lyu2024automatic}
M.~R. Lyu, B.~Ray, A.~Roychoudhury, S.~H. Tan, and P.~Thongtanunam, ``Automatic programming: Large language models and beyond,'' {\em ACM Trans. Softw. Eng. Methodol.}, Dec. 2024.

\end{thebibliography}

\newpage
\begin{IEEEbiography}{Ranim Khojah}{\,}is a PhD candidate at Chalmers University of Technology and the University of Gothenburg (Sweden). Her research interests include human-chatbot interactions in software engineering. Khojah received her Licentiate of Philosophy degree in Computer Science and Engineering from Chalmers University of Technology. Contact her at \url{khojah@chalmers.se} or \url{www.ranimkhojah.com}.
\vadjust{\vfill}
\end{IEEEbiography}

\begin{IEEEbiography}{Mazen Mohamad}{\,}is a researcher at RISE, the Research Institutes of Sweden and a lecturer at Chalmers University of Technology. He holds a Ph.D. degree in software engineering from the University of Gothenburg. His research focuses on security assurance, combined safety and security analysis, AI in software engineering, and AI for cybersecurity. Contact him at \url{mazen.mohamad@ri.se} or \url{www.mazenm.com}. 
\vadjust{\vfill}
\end{IEEEbiography}

\begin{IEEEbiography}{Linda Erlenhov}{\,}is a lecturer at Chalmers University of Technology and the University of Gothenburg in Gothenburg Sweden. Her research interests includes human aspects of software engineering and software development tooling. Erlenhov received her Licentiate of Engineering degree in Computer Science and Engineering from Chalmers University of Technology. Contact her at \url{linda.erlenhov@chalmers.se}.
\vadjust{\vfill}
\end{IEEEbiography}

\begin{IEEEbiography}{Francisco Gomes de Oliveira Neto}{\,}is an Associate Professor in Software Engineering at the University of Gothenburg as well as the Chalmers University of Technology, Sweden. His main research areas are automated software testing, and (AI) bots to aid software engineers. He received his PhD in Computer Science from the Universidade Federal de Campina Grande (UFCG, Brazil). Contact him at \url{francisco.gomes@cse.gu.se}.
\vadjust{\vfill}
\end{IEEEbiography}

\begin{IEEEbiography}{Philipp Leitner}{\,}is an Associate Professor of Software Engineering at Chalmers University of Technology as well as the University of Gothenburg (Sweden). His research interests are in empirical software engineering, with a focus on software performance optimization and the development of web- and cloud-based systems. Leitner received his doctoral degree in business informatics from TU Vienna (Austria). He is a member of the ACM. Contact him at \url{philipp.leitner@chalmers.se} or \url{https://icet-lab.eu}.
\vadjust{\vfill}
\end{IEEEbiography}

\end{document}